\documentclass[aps,prd,preprintnumbers,nofootinbib,floatfix,11pt]{revtex4}

\usepackage{amsfonts,amsmath,amssymb,mathrsfs,graphicx,epsfig,color} 
\usepackage{hyperref}

\allowdisplaybreaks[1]
\newcommand{\be}{\begin{equation}}
\newcommand{\ee}{\end{equation}}
\newcommand{\ba}{\begin{eqnarray}}
\newcommand{\ea}{\end{eqnarray}}

\begin{document}

\title{On uniqueness of static spacetime with conformal scalar in higher dimensions}
\author{Keisuke Izumi$^{1,2}$, Yoshimune Tomikawa$^{3}$ and Tetsuya Shiromizu$^{2,1}$}

\affiliation{$^{1}$Kobayashi-Maskawa Institute, Nagoya University, Nagoya 464-8602, Japan} 
\affiliation{$^{2}$Department of Mathematics, Nagoya University, Nagoya 464-8602, Japan}
\affiliation{$^{3}$Faculty of Economics, Matsuyama University, Matsuyama 790-8578, Japan}

\begin{abstract}
\begin{center}
{\bf Abstract}
\end{center}
\noindent
We discuss the uniqueness of asymptotically flat and static spacetimes in the $n$-dimensional 
Einstein-conformal scalar system. 
This theory potentially has a singular point in the field 
equations where the effective Newton constant diverges. We will show that the static spacetime 
with the conformal scalar field outside a certain surface $S_p$ associated with the singular 
point is unique. 
\end{abstract}

\maketitle

%
%
%
\section{Introduction}

The Einstein-conformal system is regarded as a toy model of 
a low-energy effective theory of string theory. 
It is known that there is a static, spherically symmetric and asymptotically flat exact solution in the $n$-dimensional 
Einstein-conformal scalar system \cite{BBMB,HDBBMB}. 
For $n=4$, the curvature is finite at the Killing horizon and it is black hole solution 
which is known as the Bocharova-Bronnikov-Melnikov-Bekenstein (BBMB) black hole \cite{BBMB}, 
whereas for $n>4$, the solution is not black hole because the 
curvature diverges at the Killing horizon \cite{HDBBMB}. 
The field equation has a singular point where the effective Newton constant diverges.
It appears in the exact solutions (the corresponding surface will be denoted by $S_p$ later) 
but all the physical quantities such as the curvature and the scalar field are regular there. 
For $n=4$, it is natural to ask if the uniqueness of the 
black hole solution holds. 
The current authors addressed this issue 
and then proved the uniqueness of the static spacetime outside $S_p$ \cite{Tomikawa2017a, Tomikawa2017b, Shinohara}. 
Since the proof in Refs. 
\cite{Tomikawa2017a, Shinohara} relies on the Gauss-Bonnet theorem, as black hole cases \cite{Israel,Robinson}, 
it works only for four dimensions. 
By contrast, the proof 
based on the conformal transformation and positive mass theorem in Ref. \cite{Tomikawa2017b} may be able to work in higher 
dimensions too as black hole cases \cite{bm, GIS, IIS}. 

In this paper, we will address the uniqueness of the static, asymptotically flat exact solution in the 
Einstein-conformal scalar system. 
The exact solution for $n>4$ has the curvature singularity on the Killing horizon, but 
it does not appear in the region between $S_p$ and the spatial infinity. 
Therefore, the uniqueness of the solution outside $S_p$ would be worth investigating.
As a consequence, following Refs. \cite{GIS, IIS}, we shall prove the uniqueness of 
static spacetime outside $S_p$. 

The rest of this paper is organized as follows. In Sec. II, we will present the setup, describe the 
basic features of static spacetime and discuss the 
boundary conditions. In Sec. III, we show that a certain relation 
between the lapse function and scalar field holds. The proof will be shown in Sec. IV. 
Finally, we will give the summary and discussion in Sec. V.

%
\section{setup}

In this section, we describe the Einstein-conformal scalar system in $n$-dimensions and discuss 
some feature of the static cases. We suppose that $n$ is larger than four. The case for $n=4$ has been investigated in  Refs. \cite{Tomikawa2017a, Tomikawa2017b}. 

We begin with the action for the Einstein-conformal scalar system in $n$-dimensions,
\begin{eqnarray}
S=\int \left[\frac{1}{2\kappa}R-\frac{1}{2}(\nabla \phi)^2-\frac{\xi}{2}R \phi^2    \right]{\sqrt {-g}}d^nx, 
\end{eqnarray}
where 
\begin{eqnarray}
\xi:=\frac{1}{4}\frac{n-2}{n-1}.
\end{eqnarray}
One can then obtain the field equations,
\begin{eqnarray}
(1-\kappa \xi \phi^2) G_{\mu\nu}=\kappa \left[ \nabla_\mu \phi \nabla_\nu \phi-\frac{1}{2}g_{\mu\nu}(\nabla \phi)^2
+\xi (g_{\mu\nu}\nabla^2-\nabla_\mu \nabla_\nu )\phi^2   \right], \label{einsteineq1}
\end{eqnarray}
where $G_{\mu \nu}$ is the $n$-dimensional Einstein tensor,
and
\begin{eqnarray}
(\nabla^2 -\xi R)\phi=0. \label{scalareq1}
\end{eqnarray}
The trace of Eq. (\ref{einsteineq1}) and Eq. (\ref{scalareq1}) show us 
\begin{eqnarray}
R=0.
\end{eqnarray}
This simplifies the field equations as
\begin{eqnarray}
(1-\kappa \xi \phi^2) R_{\mu\nu}=\kappa \left[ \nabla_\mu \phi \nabla_\nu \phi-\frac{1}{2}g_{\mu\nu}(\nabla \phi)^2
+\xi (g_{\mu\nu}\nabla^2-\nabla_\mu \nabla_\nu )\phi^2   \right] \label{einsteineq2}
\end{eqnarray}
and
\begin{eqnarray}
\nabla^2 \phi=0. \label{scalareq2}
\end{eqnarray}
Note that the front factor of the left-hand side in Eq. (\ref{einsteineq2}) tells us that the 
equation is singular at $\phi=\phi_p:=\pm 1/ {\sqrt {\kappa \xi}}$.
In the exact solutions found in Ref. \cite{HDBBMB}, however, the spacetime is regular at $S_p$ 
which denotes surface(s) satisfying $\phi=\phi_p$,
although it is singular on the Killing horizon for higher dimensions than four, 
which is different from the case of four dimension \cite{BBMB}. 

From now on, we will focus on static and asymptotically flat spacetimes. The metric can be written as 
\begin{eqnarray}
ds^2=-V^2(x^i)dt^2+g_{ij}(x^k)dx^idx^j.
\end{eqnarray}
The Einstein equation (\ref{einsteineq2}) give us 
\begin{equation}
(1- \kappa \xi \phi^2)D^2V=\kappa \left[\left( \frac{1}{2}-2\xi \right)V(D\phi)^2 +2\xi \phi D^i V D_i \phi \right] \label{einsteineq00}
\end{equation}
and
\begin{equation}
(1- \kappa \xi \phi^2) \left( {}^{(n-1)}R_{ij}-\frac{1}{V}D_iD_jV  \right)=\kappa \left[(1-2\xi)D_i \phi D_j \phi
-\left( \frac{1}{2}-2\xi\right)(D\phi)^2g_{ij}-2\xi \phi D_iD_j \phi \right], \label{einsteineqij}
\end{equation}
where $D_i$ is the covariant derivative with respect to $g_{ij}$ and ${}^{(n-1)}R_{ij}$ is the 
$(n-1)$-dimensional Ricci tensor.  
The equation for the scalar field becomes
\begin{equation}
D^i(VD_i \phi)=0.\label{scalareq3}
\end{equation}
We impose the asymptotic boundary conditions at spatial infinity ($r\to \infty$),  
\begin{eqnarray}
& & V=1-\frac{m}{r^{n-3}}+O(1/r^{n-2}),\\
& & g_{ij}=\left(1+\frac{2}{n-3}\frac{m}{r^{n-3}}\right)\delta_{ij}+O(1/r^{n-2})
\end{eqnarray}
and $\phi=O(1/r^{n-3})$. 
Equation (\ref{scalareq3}) tells us that $\phi$ is a monotonic function,
and thus $\phi$ is regular in the region whose boundaries are composed of the surface $S_p$ and the spatial infinity, 
while the Einstein equation is singular at $S_p$. 
Therefore, we shall consider only the region $\Sigma$ enclosed by $S_p$ and the spatial infinity. 
Here, we consider the case where $\phi$ has the same sign everywhere on $S_p$, i.e. $\phi_p = 1/ {\sqrt {\kappa \xi}}$.

%
\section{$\phi$-$V$ relation and foliation on $\Sigma$}

In this section, we will show the relation between the conformal scalar field $\phi$ and $V$. 
The relation gives us a harmonic function $v$ on $\Sigma$ and 
the foliation on $\Sigma$ is taken by $v$. 
We derive some formulae for geometrical quantities expressed with the $(n-2)$-dimensional ones, 
which are useful for the proof of the uniqueness of static spacetime outside the surface 
$S_p$. 

The combination of Eqs. (\ref{einsteineq00}) and (\ref{scalareq3}) gives 
\begin{equation}
D^i \left[(1-\varphi)(1+\varphi)^{\frac{n-4}{n-2}}D_i \left\lbrace  (1+\varphi)^{\frac{2}{n-2}}V  \right\rbrace   \right]=0,
\label{varphieq}
\end{equation}
where $\varphi:= {\sqrt {\kappa \xi}}\phi $. The integration over $\Sigma$ gives us 
\begin{equation}
0 =  \int_\Sigma D^i \left[(1-\varphi)(1+\varphi)^{\frac{n-4}{n-2}}D_i \left\lbrace  (1+\varphi)^{\frac{2}{n-2}}V  \right\rbrace   \right]
d \Sigma =\int_{S_\infty}D_i \left\lbrace  (1+\varphi)^{\frac{2}{n-2}}V  \right\rbrace dS^i. \label{integral1}
\end{equation}
In the second equality, we used the Gauss theorem and the fact that the boundary term from $S_p$ vanishes because of $\varphi |_{S_p}=1$. 
Next, we multiply  $(1+\varphi)^{\frac{2}{n-2}}V$ to Eq. (\ref{varphieq}) and then take its integration over $\Sigma$. 
Finally, under the current boundary conditions, we have  
\begin{eqnarray}
0&=& \int_\Sigma  \left\lbrace  (1+\varphi)^{\frac{2}{n-2}}V  \right\rbrace D^i \left((1-\varphi)(1+\varphi)^{\frac{n-4}{n-2}}\left[ D_i \left\lbrace  (1+\varphi)^{\frac{2}{n-2}}V  \right\rbrace \right]\right) d \Sigma \nonumber \\
&=&\int_{S_\infty}  \left\lbrace  (1+\varphi)^{\frac{2}{n-2}}V  \right\rbrace (1-\varphi)(1+\varphi)^{\frac{n-4}{n-2}}\left[ D_i \left\lbrace  (1+\varphi)^{\frac{2}{n-2}}V  \right\rbrace \right] d S^i \nonumber \\
&& \hspace{2mm}
-\int_\Sigma (1-\varphi)(1+\varphi)^{\frac{n-4}{n-2}}\left[ D \left\lbrace  (1+\varphi)^{\frac{2}{n-2}}V  \right\rbrace \right]^2d \Sigma, \label{integral2}
\end{eqnarray}
where we arranged the integration in the same manner as Eq. (\ref{integral1}). 
The surface intergal term over $S_\infty$ in Eq. (\ref{integral2}) has the same value as 
that in Eq. (\ref{integral1}) because of $V\to 1$ and $\varphi \to 0$ at infinity. 
This means that it is zero, and then the volume integral term over $\Sigma$ in the last line of Eq. (\ref{integral2}) vanishes. 
It occurs only if $D_i \left\lbrace  (1+\varphi)^{\frac{2}{n-2}}V  \right\rbrace =0$ holds, 
i.e., $(1+\varphi)^{\frac{2}{n-2}}V$ is constant. 
The constant value can be fixed by checking its behavior at infinity and 
we can see $(1+\varphi)^{\frac{2}{n-2}}V=1$. 
For convenience, we rearrange the relation as 
\begin{equation}
\varphi=V^{-\frac{n-2}{2}}-1. \label{relation}
\end{equation}
This relation significantly makes analysis simple. 
Using this relation, Eq. (\ref{einsteineq00}) and the trace of Eq. (\ref{einsteineqij}) imply 
\begin{equation}
D^2V=\frac{n-2}{2}\frac{(DV)^2}{V} \label{D2V}
\end{equation}
and
\begin{equation}
{}^{(n-1)}R=(n-2)\frac{(DV)^2}{V^2}. \label{n-1dimRicci}
\end{equation}
For $n \geq 5$, we define $v$ as $v:=V^{-\frac{n-4}{2}}$, and then Eq. (\ref{D2V}) becomes a simple harmonic equation for $v$
\begin{equation}
D^2v=0.
\end{equation}
This leads to the fact that $v$ is a monotonically decreasing function toward the spatial infinity.\footnote{For the $n=4$ case, 
$v$ is defined by $v:=\ln V$. See Refs. \cite{Tomikawa2017a, Tomikawa2017b}.}
Bearing this in mind, we introduce the unit normal vector to $v=$constant surface $S_v$ in $\Sigma$ as 
\begin{eqnarray}
n_i=-\rho D_i v,
\end{eqnarray}
where $\rho:=(D^ivD_iv)^{-1/2}$. The extrinsic curvature, $k_{ij}$ of $S_v$, is defined by $k_{ij}=h_i^kD_k n_j$, where 
$h_{ij}:=g_{ij}-n_in_j$ is the induced metric of $S_v$. 
For later discussions, we present the trace of the extrinsic curvature
\begin{eqnarray}
k=D^in_i=-D^i(\rho D_iv)=-D^i \rho D_i v=\frac{1}{\rho}n^iD_i \rho
\end{eqnarray}
and the second derivative of $v$
\begin{eqnarray}
D_i D_j v=-\frac{1}{\rho}k_{ij}+\frac{1}{\rho^2}(n_i {\cal D}_j \rho + n_j {\cal D}_i \rho )+\frac{k}{\rho}n_in_j,
\end{eqnarray}
where ${\cal D}_i$ is the covariant derivative with respect to $h_{ij}$. 

Let us check the regularity of the curvature invariant at $S_p$. 
We express the curvature invariant by the $(n-2)$-dimensional geometrical quantities,
\begin{eqnarray}
R_{\mu\nu}R^{\mu\nu}& = & R_{00}R^{00}+R_{ij}R^{ij} \nonumber \\
&= & 
\frac{4(n-2)^2}{(n-4)^4} \frac{1}{v^4\rho^4} \nonumber \\
&&
+\frac{4(n-2)^2}{(n-4)^4} \frac{1}{v^4\rho^4} \frac{1}{\left(2v^{-\frac{n-2}{n-4}}-1\right)^2}
\Biggl[ \Biggl\lbrace
(n-3)+2v^{-\frac{n-2}{n-4}} -(n-4)v \left( 1-v^{-\frac{n-2}{n-4}} \right) \rho k
\Biggr\rbrace ^2 \nonumber \\
&&
+\Biggl\lbrace (n-4)v \left( 1-v^{-\frac{n-2}{n-4}} \right) \rho k_{ij} -h_{ij} \Biggr\rbrace ^2
+2(n-4)^2 v^2 \left( 1-v^{-\frac{n-2}{n-4}} \right) ^2 ({\cal D} \rho )^2
\Biggr].
\label{riccisquare}
\end{eqnarray}
Equation (\ref{relation}) shows that $V$ and $v$ have the following values at $S_p$:
\begin{eqnarray}
V=V_p:=2^{-\frac{2}{n-2}}~~{\rm and}~~v=v_p:=2^{\frac{n-4}{n-2}}.
\end{eqnarray}
The second one of the above indicates that the curvature invariant has a vanishing factor in a denominator, 
which naively leads to a curvature singularity. 
This singularity can be avoided only if 
\begin{eqnarray}
k_{ij}|_{S_p}=\frac{2^{\frac{2}{n-2}}}{n-4}  \frac{1}{\rho_p} h_{ij}|_{S_p}, \label{extbc}
\end{eqnarray}
where $\rho_p:= \rho |_{S_p}$, and
\begin{eqnarray}
{\cal D}_i \rho|_{S_p}=0 \label{drho}
\end{eqnarray}
hold. 
They mean that $S_p$ is totally umbilic and that $\rho$ is constant on $S_p$. 
We should also check the behavior of the Kretschmann invariant, which is decomposed as 
\begin{eqnarray}
R_{\mu\nu\rho\sigma}R^{\mu\nu\rho\sigma}=4R_{0i0j}R^{0i0j}+R_{ijkl}R^{ijkl}. \label{riemannsquare}
\end{eqnarray}
The first term in the right-hand side becomes 
\begin{eqnarray}
R_{0i0j}R^{0i0j} &=& \frac{4}{V^2}D_iD_jVD^iD^jV \nonumber \\
&=&\frac{4}{(n-4)^2} V^{n-4} \left[ 
\frac{1}{\rho^2}k_{ij}k^{ij}+\frac{2}{\rho^4}({\cal D}\rho)^2+\frac{1}{\rho^2}\left(k-\frac{n-2}{n-4}\frac{1}{\rho v}\right)^2
\right].
\end{eqnarray}
The second term in the right-hand side of Eq. (\ref{riemannsquare}) is composed of the $(n-1)$-dimensional Weyl tensor square and Ricci tensor square. 
The Ricci tensor square has the same singular behavior as Eq. (\ref{riccisquare}).
As will be seen in the next section, Eqs. (\ref{extbc}) and (\ref{drho}) are sufficient for the proof of the spherical symmetry and the uniqueness. 
After the spherical symmetry is shown, we know that the Weyl tensor vanishes. 
Therefore, the form of Weyl tensor square is not important in our proof and we do not show it explicitly. 

Note that, at the Killing horizon ($V=0$ or $v \to \infty$), the curvature invariant does not 
imply the constraint on the geometry.

%
\section{Proof for uniqueness}

Now it is ready to prove the uniqueness of static spacetime outside $S_p$ in the $n$-dimensional 
Einstein-conformal scalar system. The procedure for the proof follows Refs. \cite{GIS,IIS}. 

Let us consider the following two conformal transformations for $g_{ij}$:
\begin{eqnarray}
\tilde g_{ij}^\pm=\Omega_\pm^2 g_{ij},
\end{eqnarray}
where 
\begin{eqnarray}
& & \Omega_+=V^{\frac{1}{n-3}} \label{Omega+} \\
& & \Omega_-=\frac{1}{V}\left(1-V^{\frac{n-2}{2}} \right)^{\frac{2}{n-3}}.  \label{Omega-}
\end{eqnarray}
Then we have two manifolds $(\tilde \Sigma^\pm, \tilde g_{ij}^\pm)$. 

It is easy to see that the Ricci scalar for $\tilde g_{ij}^\pm$ vanishes,  
\begin{eqnarray}
\Omega_\pm^2{}^{(n-1)}\tilde R_\pm={}^{(n-1)}R-2(n-2)D^2 \ln \Omega_\pm-(n-3)(n-2)(D \ln \Omega_\pm)^2=0.
\end{eqnarray}
The metrics $\tilde g_{ij}^\pm$ on $S_p$ coincide with each other,
\begin{eqnarray}
\tilde g_{ij}^+|_{S_p}=\tilde g_{ij}^-|_{S_p}=2^{-\frac{4}{(n-2)(n-3)}} g_{ij}|_{S_p}.
\end{eqnarray}
The extrinsic curvature of $v=$constant surface in $\tilde \Sigma^\pm$ is related to that in 
$\Sigma$ as 
\begin{eqnarray}
\tilde k_{ij}^\pm=\Omega_\pm k_{ij}+\frac{2}{n-4}v^{-\frac{n-2}{n-4}}\frac{1}{\rho}\frac{d \Omega_\pm}{dV} h_{ij}.
\label{tildek}
\end{eqnarray}
Substituting  Eqs. (\ref{extbc}), (\ref{Omega+}) and (\ref{Omega-}) into Eq. (\ref{tildek}), we can see 
\begin{eqnarray}
\tilde k_{ij}^+|_{S_p}=-\tilde k^-_{ij}|_{S_p}=\frac{n-2}{(n-3)(n-4)}2^{\frac{2(n-4)}{(n-2)(n-3)}}\frac{1}{\rho_p}h_{ij}|_{S_p},
\label{tildek+-}
\end{eqnarray}
where $\rho_p:=\rho|_{S_p}$. 
The coincidence of the metric and the relation of the extrinsic curvature at $S_p$ allow us to glue $\tilde \Sigma^+$ 
and $\tilde\Sigma^-$ without jump of the extrinsic curvature. 
The glued manifold $\tilde \Sigma^+ \cup \tilde \Sigma^- $ is denoted by $(\tilde \Sigma, \tilde g_{ij})$.

Here we examine the asymptotic behavior of $\tilde g_{ij}^\pm$. After short calculation, we have 
\begin{eqnarray}
\tilde g_{ij}^+=\delta_{ij}+O(1/r^{n-2}).
\end{eqnarray}
This implies that the mass vanishes. For $\tilde g_{ij}^-$, by introducing the new radial coordinate $\chi$ as 
\begin{eqnarray}
\chi:=\left(\frac{n-2}{2}m\right)^{\frac{2}{n-3}}r^{-1}, 
\end{eqnarray}
the neighborhood of $r\to\infty$ ($\chi=0$) can be approximated by
\begin{eqnarray}
\tilde g_{ij}^-dx^idx^j \simeq d\chi^2+\chi^2 d\Omega_{n-2}^2,
\end{eqnarray}
where $d\Omega_{n-2}^2$ is the metric of the $(n-2)$-dimensional round sphere. This means that 
the spatial infinity in $\Sigma$ corresponds to a point in $\tilde \Sigma$.
Therefore, 
adding one point $\{q\}$ corresponding to $\chi=0$, we have a complete manifold $\tilde \Sigma \cup {q}$, 
which is asymptotically flat space with the zero mass and zero Ricci scalar. 
Now we can apply the positive mass theorem for $(\tilde \Sigma, \tilde g_{ij})$ and then 
$(\tilde \Sigma, \tilde g_{ij})$ is flat space, that is, $\tilde g_{ij}=\delta_{ij}$. This shows us 
that Eq. (\ref{D2V}) is rewritten as 
\begin{eqnarray}
\Delta_\delta V^{-\frac{n-2}{2}}=0, \label{flatlaplace}
\end{eqnarray}
where $\Delta_\delta$ is the flat Laplacian. 
Since we have already shown in Eq. (\ref{tildek+-}) that $S_p$ is totally umbilic in $\tilde \Sigma$, 
$S_p$ in $\tilde \Sigma$ is spherically symmetric because of the flatness of $\tilde \Sigma$. 
The fact that the solution to Eq. (\ref{flatlaplace}) with the spherically symmetric boundary condition 
is spherically symmetric leads to the result that every $v=$constant surface $\lbrace S_v \rbrace$ in $\tilde \Sigma^+$ is also 
spherically symmetric. Thus, $(\Sigma,g_{ij})$ is spherically symmetric because the conformal factor 
$\Omega_\pm$ depends only on $V$. 
Then the results in Ref. \cite{HDBBMB} show us that the solution is unique outside $S_p$.

%
\section{Summary and discussion}
\label{summary}

In this paper, we proved the uniqueness of static spacetime outside the surface $S_p$, where the 
field equations are singular, in the $n$-dimensional Einstein-conformal scalar system. Therefore, as 
the event horizon in the vacuum Einstein, the regularity on the surface $S_p$ could constrain the 
geometry. This is the crucial first step to prove the uniqueness. In the end, we could prove it. 

Our uniqueness theorem says nothing for whole spacetime. In contrast to the four dimensional cases, 
the exact solution of static, spherically symmetric spacetime is singular at the Killing horizon \cite{HDBBMB}.
It would be difficult to resolve the singularities and obtain 
any constraints for the geometry of the Killing horizon. 
Yet, even if some geometrical quantities such as the extrinsic curvature diverge at the 
Killing horizon, the curvature invariant may be finite. This is rather impressive because all 
geometrical quantities at the 
Killing horizon are finite in four dimensions.

%

\acknowledgments

K. I. and T. S. are supported by Grant-Aid for Scientific Research from Ministry of Education, 
Science, Sports and Culture of Japan (No. JP17H01091). K.~I. is also supported by 
JSPS Grants-in-Aid for Scientific Research (B) (JP20H01902). T. S. is also supported by 
JSPS Grants-in-Aid for Scientific Research (C) (JP21K03551).


%
%

\end{document}